\documentclass[aps,twocolumn,nofootinbib,preprintnumbers,prl,10pt]{revtex4-2}
\usepackage[utf8]{inputenc}

\usepackage[svgnames]{xcolor}
\usepackage[most]{tcolorbox}
\usepackage{caption}
\tcbset{colback=PaleGreen!0!white, colframe=NavyBlue, 
	highlight math style= {enhanced, 
		colframe=red,colback=red!10!white,boxsep=0pt}
}
\usepackage{relsize}

\usepackage{amsmath}
\usepackage[bottom]{footmisc}
\usepackage{braket}
\usepackage{graphicx}
\usepackage{mwe}
\usepackage[section]{placeins}
\usepackage{framed}
\usepackage{csquotes}
\usepackage{tikz}
\usetikzlibrary{decorations.markings}
\usetikzlibrary{decorations.pathmorphing}
\definecolor{shadecolor}{rgb}{0.90,0.90,0.90}
\usepackage{hyperref}
\usepackage{bbold}
\usepackage{subcaption}
\usepackage{pifont}
\usepackage{setspace}
\usepackage{amsmath, amssymb, amsthm, float, graphicx,amsfonts}
\usepackage{amsthm}
\usepackage{array, boldline, makecell, booktabs}

\theoremstyle{definition}

\hypersetup{colorlinks=true, linkcolor=DarkRed, citecolor=DarkRed,urlcolor=Indigo,linktocpage}
\def\beq{\begin{eqnarray}}\def\eeq{\end{eqnarray}}
\def\be{\begin{equation}}\def\ee{\end{equation}}
\def\bs{\begin{split}}\def\es{\end{split}}

\usepackage{bm}

\def \zbr{\bar{z}}

\usepackage{mathrsfs}
\usepackage{slashed}
\newcommand{\scrip}{\mathscr{I}^{+}}
\newcommand{\scrim}{\mathscr{I}^{-}}

\begin{document}

\title{\bf Hypertranslations and Hyperrotations}

\author{\!\!\!\! Chethan Krishnan$^{a}$ and Jude Pereira$^{b}$\\ ~~~~\\
\it ${^a}$Centre for High Energy Physics,
\it Indian Institute of Science,\\ \it C.V. Raman Road, Bangalore 560012, India. \\
{\rm Email:} \textmd{chethan.krishnan@gmail.com}\\
\it ${^b}$ Department of Physics, Arizona State University,\\ Tempe, Arizona 85287-1504, USA. \\
{\rm Email:} \textmd{jude.pereira@asu.edu}}

\begin{abstract}{We study the asymptotic symmetries of Einstein gravity in flat space. Instead of Bondi gauge, we work with the recently introduced special double null gauge, in which $\scrip$ and $\scrim$ are approached along null directions. We find four new functions worth of asymptotic diffeomorphisms beyond the familiar supertranslations and superrotations, which are of relevance in discussions of finite surface charges. Two of these arise from angle-dependent shifts in the $v$-coordinate near $\scrip$. We call these hypertranslations and sub-leading hypertranslations, with analogous statements in the $u$-coordinate near $\scrim$. There are also two Diff$(S^2)$ transformations, which we call hyperrotations, that are sub-leading to the Virasoro superrotations. With power law fall-offs in the null coordinate and the standard metric on the sphere at leading order, we prove that this is the exhaustive list of diffeomorphisms whose associated metric parameters can show up in the (finite) surface charges. We compute the algebra of the asymptotic Killing vectors under the Barnich-Troessaert bracket, and find a four-fold infinite generalization of the BMS algebra. 


 
}
\end{abstract}
\maketitle

{\noindent \bf Introduction}: To formulate quantum gravity holographically in flat space, an understanding of the asymptotic structure and symmetries will likely be essential.  Attempts in this direction typically consider the Bondi gauge \cite{Bondi} and future null infinity $\scrip$, or the Ashtekar-Hansen gauge \cite{Ashtekar} and spatial infinity $i^0$. Based on various motivations\footnote{See eg. discussions in \cite{KP, CK, KPP, KPBig}.}, a new gauge for asymptotically flat space was introduced recently in \cite{KP} which treats $\scrip$ and $\scrim$ on an equal footing, and where these boundaries are approached along null coordinates $v$ and $u$. With a specific choice of fall-offs in the null coordinates, it was found that this Special Double Null (SDN) gauge \cite{KP} can reproduce the famous BMS algebra. Some hints that a generalization of BMS may be accessible in this gauge were noted, but because of various novel technical and conceptual features that emerged in the gauge which needed immediate attention \cite{KP, KPBig}, a detailed exploration of this was not undertaken in \cite{KP}.

In this paper, we remedy this situation. We write down the complete symmetry algebra of asymptotic diffeomorphisms in the SDN gauge that can be relevant in discussions of finite Iyer-Wald-Barnich-Brandt \cite{Iyer, Brandt} surface charges. We do this under two assumptions: (a) the leading piece of the metric on the sphere is of the standard form, and (b) the fall-offs in the null coordinates are power law\footnote{This may seem overly restrictive for the following reason: the double null form of the Schwarzschild metric that one usually writes down (see eg. eqn (24) of \cite{KP}), contains log fall-offs. But it is possible to write down Schwarzschild in the SDN gauge, without log fall-offs \cite{KPBig}. 
}. We identify  fall-offs in the metric that are preserved by these asymptotic diffeomorphisms. The result is that there is an enhancement of the BMS algebra beyond the supertranslations and superrotations to include four more functions on the sphere. Two of these arise from angle-dependent shifts at $O\big(v^0\big)$ and $O\big(v^{-1}\big)$ in the $v$-coordinate at $\scrip$. Since the supertranslations arise from shifts in $u$ at $\scrip$, we call these {\em hypertranslations}\footnote{Hypertranslations and sub-leading hypertranslations, when we need to distinguish the order of $v$-fall-off at which they appear.} by analogy. There are also two non-holomorphic diffeomorphisms of the sphere which are sub-leading to the (super)rotations. We call these {\em hyperrotations}. Overall this leads to the Beyond-BMS (BBMS) algebra being a 4-fold infinite generalization of the usual BMS algebra. We have checked that in the finite surface charges, only those metric parameters or gauge parameters corresponding to the BBMS diffeomorphisms (and none of the  ``trivial'' diffeomorphisms noted in \cite{KP} that arise at further subleading orders) appear. This is shown in the Supplementary Material.  After the antipodal mapping between $\scrip_-$ and $\scrim_+$ is imposed \cite{Strominger, KP}\footnote{In the SDN gauge, this has a direct understanding as due to asymptotic CPT invariance \cite{KP}.}, the two BBMS algebras at $\scrip$ and $\scrim$ collapse to a single diagonal BBMS$^0$ algebra.

We emphasize that the results we find in this paper are obtained via an entirely conventional study of asymptotic symmetries of Einstein gravity in flat space. The new ingredient is that we are working with a new gauge, which we believe is better motivated for holography \cite{KP}. So our results on {\em asymptotic} symmetries should be distinguished from the recent discussions on {\em celestial} symmetries, which are identified at the level of scattering amplitudes of massless particles expressed in the celestial basis (see eg. \cite{Pasterski}). The connection between celestial and asymptotic symmetries is not entirely transparent, but they are believed to be indicators of the same physics. Some structural distinctions between conventional BMS transformations and the extra BBMS ones we find here will be discussed in the Discussion section and the Supplementary Material. But a detailed study of this is clearly necessary.








{\noindent \bf{Asymptotic Killing Vectors}}: The gauge we work with was discussed in \cite{KP} and the fall-offs we will consider are presented in the Supplementary Material. For technical reasons, it will be useful for us to write the fall-offs in terms of upstairs indices, where they can be written as
\begin{subequations}\label{boundarycondnew}
    \begin{align}
        g^{uv} &= -2+O\big(v^{-2}\big)\\
        g^{AB} &= 4\gamma^{AB}\, v^{-2}+O\big(v^{-3}\big)\\
        g^{uA} &= g^{vA} = O\big(v^{-3}\big)
    \end{align}
\end{subequations}
Corresponding to these metric fall-offs, we will write a set of asymptotic Killing vector conditions as follows:
\begin{subequations}\label{approxLie}
    \begin{align}
        \label{approxLieuv}
        \mathcal{L}_{\xi}g^{uv} &= O\big(v^{-2}\big)\\
        \label{approxLieuA}
        \mathcal{L}_{\xi}g^{uA} &= O\big(v^{-3}\big) \\
        \label{approxLievA}
        \mathcal{L}_{\xi}g^{vA} &= O\big(v^{-3}\big) \\
        \label{approxLieAB}
        \mathcal{L}_{\xi}g^{AB} &= O\big(v^{-3}\big)
    \end{align}
\end{subequations}
Solving these along with the exact Lie derivative conditions \eqref{exactKilling}, we obtain the following form of the asymptotic Killing vectors
\begin{subequations}\label{finalxi}
    \begin{align}
        \xi^u &= f+ \frac{1}{2}\, \alpha^A_3\, \partial_A f\, v^{-2} + \frac{1}{3}\, \alpha^A_4\, \partial_A f \, v^{-3}+O\big(v^{-4}\big) \\
        \xi^v &= -\frac{\psi}{2}\, v+\xi^v_{(0)}\,+\frac{\xi^v_{(1)}\,}{v}+O\big(v^{-2}\big) \\
        \xi^A &= Y^A-2\, \gamma^{AB}\, \partial_{B} f\, v^{-1}+\xi^A_{(2)}\, v^{-2}+O\big(v^{-3}\big)      
    \end{align}
\end{subequations}
where \beq\label{fdef} f=\xi_{0}^{u}=\psi(z,\zbr)\, u/2 + T(z,\zbr),  \ \ {\rm with} \ \ \psi(z,\zbr)= D_AY^A. \nonumber \\\eeq
Note that we have used $T(z,\zbr)$ to denote supertranslations and $Y^{z}(z), Y^{\zbr}(\zbr)$ to denote superrotations, as in \cite{KP}. 

In addition to these familiar BMS diffeomorphisms, the exact and asymptotic Killing vector equations also determine $\xi^v_{(0)}, \xi^v_{(1)}$ and $\xi^A_{(2)}$. They contain the independent functions that we call \emph{hypertranslations} $\phi(z,\zbr)$,  \emph{sub-leading hypertranslations} $\tau(z,\zbr)$, and {\em hyperrotations} $Z^{A}(z,\zbr)$ respectively. The precise relation between the $\xi^v_{(0)}, \xi^v_{(1)}, \xi^A_{(2)}$ and these functions, is as follows:
\begin{subequations}\label{xidef}
    \begin{align}
\xi^v_{(0)}&=\phi+ T+\triangle_\gamma T, \\
\xi^v_{(1)}&= \tau -\frac{1}{4}\, a^A\, \partial_A\psi\ + \nonumber\\
&+\big(D_{\zbr}\mathcal{C}^{z\zbr}_{(0)}-D_z\mathcal{C}^{zz}_{(0)}\big)\, D_zT+\big(D_{z}\mathcal{C}^{z\zbr}_{(0)}-D_{\zbr}\mathcal{C}^{\zbr\zbr}_{(0)}\big)\, D_{\zbr}T, \nonumber \\
\\
\xi_{(2)}^A &= \tilde Z^{A}+\mathscr{C}^{AB}\, D_B\psi- u^2\, D^A\psi+2\, u\, D^A\xi^v_{(0)} \nonumber \\
\end{align}
\end{subequations}
Here for convenience of presentation we have introduced  
\begin{eqnarray}
\tilde Z^{z}\equiv Z^z+\mathcal{C}^{zz}_{(0)}\, D_zT+\mathcal{C}^{z\zbr}_{(0)}\, D_{\zbr}T+T D_z\mathcal{C}^{zz}_{(0)}-T D_{\zbr}\mathcal{C}^{z\zbr}_{(0)} \nonumber \\
\label{Z1}\\
\tilde Z^{\zbr}\equiv Z^{\zbr}+\mathcal{C}^{\zbr\zbr}_{(0)}\, D_{\zbr}T+\mathcal{C}^{\zbr z}_{(0)}\, D_{z}T+T D_{\zbr}\mathcal{C}^{\zbr\zbr}_{(0)}-T D_{z}\mathcal{C}^{z\zbr}_{(0)}\nonumber \\ \label{Z2}
\end{eqnarray}
with the untilded $Z^{A}$ being the hyperrotations. 

Some explanations are required. We have defined the integration ``constants'' in the shear via\footnote{The placement of $(0)$ in the integration ``constant'' as a subscript or superscript, is not important and is done as is convenient.}
\beq \mathcal{C}_{AB}(u,z,\zbr)=\mathcal{C}^{(0)}_{AB}(z,\zbr)+\int_{-\infty}^{u}du'\mathcal{N}_{AB}(u',z,\zbr)
\eeq
where the double null news tensor $\mathcal{N}_{AB}\equiv \partial_u\mathcal{C}_{AB}$. Similarly, the integration ``constant'' $a^A$ in $\alpha^A_3$ is defined in \eqref{alpha3z} in the Supplementary Material. In this paper such integrals from $\scrip_-$ will be assumed to be well-defined, either via Christodoulou-Klainerman-like fall-off demands \cite{CK1} or (perhaps more preferably from a holographic perspective) via suitable renormalization of the integration ``constant'' after regulating the lower limit of the integral \cite{KP, KPBig}. Let us also note that the Einstein equations force $\mathcal{N}_{z\zbr}=0$, so it should be kept in mind that on-shell,
\beq
\mathcal{C}^{z\zbr}(u,z,\zbr)=\mathcal{C}^{z\zbr}_{(0)}(z,\zbr).
\eeq

The $\triangle_\gamma$ is simply notation: it stands for the Laplace operator of the standard 2-sphere metric \eqref{complexsphere}. The key point about \eqref{xidef} is that it is convenient to extract some pieces from the $\xi^v_{(i)}$'s to define the most natural notions of (leading and subleading) hypertranslations. There are multiple reasons why this is natural (and indeed necessary) as we will explain in the Supplementary Material. For the moment, this can be viewed as merely a convenient shift in their definitions. 

A similar (but not quite identical) statement applies also to $\xi_{(2)}^A$. By combining all the relevant exact and approximate Killing vector conditions, we can write \cite{KPBig} the  constraint on $\xi^A_{(2)}$ as
\beq\label{xi2A}\begin{aligned}
\partial_u\xi_{(2)}^A &= \mathcal{C}^{AB}\, D_B\psi-2\, u\, D^A\psi+2\, D^A\xi^v_{(0)} \\\implies \xi_{(2)}^A &= \mathscr{C}^{AB}\, D_B\psi- u^2\, D^A\psi+2\, u\, D^A\xi^v_{(0)}  +\tilde Z^{A}(z,\zbr)
\end{aligned}\eeq
In the second line, we have simply integrated the equation in the first line. Note that in doing this, the $u$-independence of $\psi$ and (from \eqref{xidef}) $\xi^v_{(0)}$, plays a role. Also $\mathscr{C}^{AB}$ can be thought of as being defined via
\begin{eqnarray}
\mathscr{C}^{AB}=
\int^{u}\mathcal{C}^{AB}(u,z, \zbr) du \label{integral}
\end{eqnarray}
where we have absorbed the integration ``constant" into $\tilde Z$. For our present purposes, the only point to be taken from \eqref{xi2A} is that $\tilde Z^{A}(z,\zbr)$ is identified as the $u$-independent piece in $\xi^A_{(2)}$ by absorbing all the integration ``constants'' suitably. It is this $\tilde Z^A$ that gets shifted according to \eqref{Z1}-\eqref{Z2} to obtain the hyperrotations $Z^A$. This shift is the analogue of the shift in $\xi^v_{(i)}$ that we discussed in the previous paragraph to define $\phi$ and $\tau$.

For future use, let us note that the hypertranslations $\phi(z,\zbr)$ are associated with $u$-independent shifts of the metric coefficient function $\mathcal{C}_{z\zbr}$,  the subleading hypertranslations $\tau(z,\zbr)$ are associated with $u$-independent shifts of $\lambda_2$, while hyperrotations $Z^{A}(z,\zbr)$ are associated with $u$-independent shifts of $\alpha_{3}^{\ A}$. This will be discussed in the Supplementary Material, which will also be used to justify the shifts involved in the definition of $\phi, \tau$ and $Z^A$ that we mentioned earlier. Note that supertranslations shift the  $\mathcal{C}_{zz}$ and $\mathcal{C}_{\zbr\zbr}$.  
Supertranslations and leading-\&-subleading hypertranslations are diffeomorphisms on $u$ and $v$ respectively, and hyperrotations are subleading to superrotations on the sphere. 
Also note that these diffeomorphisms are different from the more subleading diffeomorphisms \cite{KP}, in that they and/or the metric parameters affected by them, appear in the expressions for the covariant surface charges as we will also show.\\

{\bf \noindent Beyond-BMS (BBMS) Algebra}: We define the Beyond BMS algebra $\mathfrak{b}$-$\mathfrak{bms}_4$ as the asymptotic symmetry algebra formed by the seven non-trivial diffeomorphisms -- namely supertranslations, superrotations, hypertranslations \& subleading hypertranslations, and hyperrotations. We will define the bracket adapting the notation in \cite{Barnich}:
\begin{widetext}
\begin{eqnarray}
\label{YTtaubracket}\big(\widehat{Y},\widehat{T},\widehat{\phi},\widehat{\tau},\widehat{Z}\big)=\big[(Y_1,T_1,
\phi_1,\tau_1,Z_1),(Y_2,T_2,\tau_2,\phi_2,Z_2)\big]
\end{eqnarray}
\end{widetext}
The list of symbols in a pair of parentheses stands for the AKV defined by those functions. In parallel to the BMS discussion in \cite{Barnich}, we will define the commutator by determining how the capped quantities on the left hand side are defined in terms of the two sets of uncapped quantities on the right. In the SDN gauge, $\widehat{Y}$ and $\widehat{T}$ were determined to be \cite{KP}
\begin{subequations}\label{YhatThat}
    \begin{align}
        \widehat{Y}^A &= Y_1^B\, \partial_B Y_2^A-Y_2^B\, \partial_B Y_1^A \\
        \widehat{T} &= Y_1^A\, \partial_A T_2-Y_2^A\, \partial_A T_1+\frac{1}{2}\, (T_1\, \psi_2-T_2\, \psi_1).
    \end{align}
\end{subequations}
This matched the result obtained in the Bondi gauge in \cite{Barnich} thereby reproducing the BMS algebra in the SDN gauge. In addition to this, in this paper, we find that $\widehat{\phi}$, $\widehat{\tau}$ and $\widehat{Z}$ are given by the following expressions:
\begin{widetext}
\begin{subequations}\label{allhats}
    \begin{align}\label{phihat}
    \widehat{\phi} &=\frac{1}{2}(\psi_1\phi_2-\psi_2\phi_1)+\big(Y_{1}^{A}\partial_A\phi_2-Y_{2}^{A}\partial_A\phi_1\big)\\
    \begin{split}\label{tauhat}
    \widehat{\tau} &= (\psi_1\tau_2-\psi_2\tau_1)+\big(Y_1^A\partial_A\tau_2-Y_2^A\partial_A\tau_1\big)
    \end{split}\\
    \begin{split}\label{Zhat}
    \widehat{Z}^A &= \big(\psi_{1}Z^A_{2}-\psi_{2}Z^A_{1}\big)+\big(Y_1^{B}\partial_B Z^A_2-Y_2^{B}\partial_B Z^A_1\big)+\big(Z^B_1\partial_BY_2^{A}-Z^B_2\partial_BY_1^{A}\big)
    \end{split}
    \end{align}
\end{subequations}
\end{widetext}

The fact that these seven non-trivial diffeomorphisms form a closed algebra is checked by first defining the Barnich-Troessaert bracket as in \cite{Barnich}
\beq\label{modLiebracket}[\xi_1,\xi_2]_M = [\xi_1,\xi_2] - \delta^g_{\xi_1}\xi_2 + \delta^g_{\xi_2}\xi_1\eeq
where $\delta^g_{\xi_1}\xi_2$ is used to denote the variation of $\xi_2$ due to variation of the metric induced by $\xi_1$.
Using these we can check by direct calculation that the exact Killing conditions hold for the Barnich-Troessaert bracket of the two AKVs $\xi_1$ and $\xi_2$. In components, this amounts to checking that
\begin{widetext}
\begin{subequations}\label{BBMSexact}
    \begin{align}
        \label{BMSu}
        &\partial_v\big([\xi_1,\xi_2]^u_M\big)+\alpha^A\, \partial_A\big([\xi_1,\xi_2]^u_M\big) = 0\\
        \label{BMSv}
        &\partial_u\big([\xi_1,\xi_2]^v_M\big)+\alpha^A\, \partial_A\big([\xi_1,\xi_2]^v_M\big) = 0\\
        \label{BMSA}
        \begin{split}
            \partial_v\big([\xi_1,\xi_2]^A_M\big)-\partial_u\big([\xi_1,\xi_2]^A_M\big) &= -\frac{1}{2}\, \Big[\partial_B\big([\xi_1,\xi_2]^v_M\big)-\partial_B\big([\xi_1,\xi_2]^u_M\big)\Big]\, g^{BA}\, e^{\lambda}+ \\ &+ \Big[\partial_u\big([\xi_1,\xi_2]^v_M\big)-\partial_u\big([\xi_1,\xi_2]^u_M\big)+\partial_v\big([\xi_1,\xi_2]^v_M\big)-\partial_v\big([\xi_1,\xi_2]^u_M\big)\Big]\, \alpha^A
        \end{split}
    \end{align}
\end{subequations}
\end{widetext}
To identify the capped quantities we do the following expansion near $\mathscr{I}^{+}$:
\begin{subequations}\label{BBMSapprox}
    \begin{align}
        [\xi_1,\xi_2]^u_M &= \widehat{f}+O\big(v^{-1}\big) \\
        [\xi_1,\xi_2]^v_M &= -\partial_u\widehat{f}\, v+\widehat{\xi}^v_{(0)}\,+\frac{\widehat{\xi}^v_{(1)}\,}{v}+O\big(v^{-2}\big) \\
        [\xi_1,\xi_2]^A_M &= \widehat{Y}^A-2\, \gamma^{AB}\, \partial_{B} \widehat{f}\, v^{-1}+\widehat{\xi}^A_{(2)}\, v^{-2}+O\big(v^{-3}\big)      
    \end{align}
\end{subequations}
where $\widehat{f}=\big(D_A\widehat{Y}^A\big)\, u/2 + \widehat{T}$ with $\widehat{Y}^A$ and $\widehat{T}$ defined in \eqref{YhatThat} and $\widehat{\xi}^v_{(0)}, \widehat{\xi}^v_{(1)}$ and $\widehat{\xi}_{(2)}^A$ are defined as in \eqref{xidef} but with $Y^A, T, \phi, \tau, Z^A$ replaced by their capped versions, defined in \eqref{YhatThat} and \eqref{allhats}. This is a lengthy, but straightforward computation.

Together with the exact Killing conditions, the fact that the right hand side of the Barnich-Troessaert commutators of two AKVs can be written asymptotically as an AKV with the hatted diffeomorphisms, proves the validity of the algebra that we set out to prove in \eqref{YhatThat} and \eqref{allhats}. Together, \eqref{YhatThat} and \eqref{allhats} define the BBMS algebra. Note that this algebra contains the BMS algebra \eqref{YhatThat} as a subalgebra, explaining our choice of nomenclature. It is also worth emphasizing that in order to establish the BBMS algebra, we have had to keep track of higher order terms in the fall-offs. This was not necessary in the case of BMS \cite{Barnich, CK}. \\

{\noindent \bf{Discussion}}: In \cite{KP, KPBig} it was noted that Einstein equations in the SDN gauge determine (some of) the gauge functions in the metric only up to a $u$-derivative at each order, and not algebraically as in the Bondi gauge. This resulted in infinite towers of free functions of only the sphere coordinates, in the metric. Correspondingly, infinite towers of diffeomorphisms were noted in the AKVs which lead to the possibility that these functions correspond to trivial diffeomorphisms. It was observed in \cite{KP, KPBig} that for high enough orders, the parameters associated to these diffeomorphisms do not appear in the finite surface charges, proving that they are indeed trivial in Einstein gravity. The results of this paper determine the complete algebra of {\em marginal} asymptotic symmetries in the SDN gauge, under the assumption that the leading asymptotic metric is Minkowski. Since the gauge we are working with is new and has various new features \cite{KP}, it is worth clarifying what we mean by a marginal symmetry. This is done towards the end of the final section in the Supplementary Material where we discuss the asymptotic charges. We will see that the extra diffeomorphisms we found in this paper are in some ways half-way between trivial diffeomorphisms and genuine asymptotic symmetries. It will be interesting to study this further. 

We feel that an interesting open problem is to write down the algebra of the AKVs for {\em all} these infinite diffeomorphisms, without worrying about the theory (and therefore the finiteness of charges). It is noteworthy here, that demanding asymptotic Riemann flatness still leaves these functions on the sphere, unfixed \cite{KP, KPBig}. In other words, these are the asymptotic invariances of flat space, without reference to any theory. The technical challenge here is to identify the shifts that we discussed earlier, to all orders. There are some natural guesses, but a thorough discussion of this issue will be presented in future work \cite{KPInfty}. It will be interesting to see how this infinite ``symmetry'' ties up with the recent discussions \cite{StromingerW} of $w_{1+\infty}$ algebra in the celestial holography context. See \cite{Freidel} where proposals have been made for the asymptotic origin of $w_{1+\infty}$ using linearized gravity.

More broadly, it will be instructive to explore the connection between asymptotic symmetries in our gauge, and celestial symmetries. The simplest (and most striking) example of the correspondence between celestial and asymptotic symmetries is the connection \cite{Prahar} between BMS supertranslations and Weinberg's soft graviton theorem \cite{Weinberg}. We suspect that such a straightforward correspondence should exist between the hypertranslations (and perhaps hyperrotations) we have identified in this paper, and the (sub-)subleading soft graviton theorem(s) \cite{Cachazo}. The rationale behind this expectation is that hypertranslations appear in our discussion in a manner directly analogous to supertranslations, but on the ``other'' null coordinate. This is certainly worth exploration, but we will not discuss this further here. See \cite{Laddha} for a discussion of Diff($S^2$) symmetries in the Bondi gauge and their connection to soft theorems. 

Let us also note a technically trivial, but conceptually important point. Following the discussion of \cite{KP}, we can show that the symmetry of asymptotically flat space in the SDN gauge after antipodal mapping is the diagonal BBMS algebra, BBMS$^0$. There are no new ideas here on top of those encountered in the context of BMS$^0$, but it is worth emphasizing that the mapping of metric functions and gauge parameters carries through here as well, between $\scrip_-$ and $\scrim_+$. This is an indication that the BBMS symmetries are indeed symmetries of the massless S-matrix.

An interesting sub-algebra of the BBMS algebra is to set the sub-leading hypertranslations and hyperrotations to zero. This would mean that we are working with supertranslations, leading hypertranslations and superrotations. This is in many ways the simplest generalization of the BMS algebra, because it is natural to treat super and (leading) hypertranslations on an equal footing in our gauge. Hypertranslations simply turn on more components of the shear field $\mathcal{C}^{AB}$ than were accessible in the Bondi gauge. Note for example that both in $\xi^u$ and $\xi^v$, the superrotations combine with super/hypertranslations in a similar way -- to the extent that the name $\tilde T(z,\zbr)$ would have been an acceptable one for $\phi(z,\zbr)$ to emphasize its parallels with $T(z,\zbr)$.

Let us conclude by noting a minor caveat. We have not allowed $\alpha^A_2$ in our fall-offs. This is because this results in the asymptotic behavior $g_{uA}=g_{vA}=O(v^0)$. This is {\em not} subleading to Minkowski space, which has $g_{uA}=g_{vA}=0$. But despite this, it is worth pointing out that allowing $\alpha^A_2$ still leads to finite well-defined charges as can be checked by direct calculation -- we will suppress the details. If one allows $\alpha^A_2$, we find a corresponding set of Diff($S^2$) transformations that appear at $O(v^{-1})$ in $\xi^A$. These are subleading to the superrotations, but they appear earlier than the hyperrotations we have discussed in the present paper. It may be of some interest to allow $\alpha^A_2$ and consider the enhanced algebra that includes these extra hyperrotations. We will not discuss it here.

\section*{Acknowledgments}
 
We thank Shamik Banerjee and Sudip Ghosh for discussions on celestial symmetries.

\vspace{0.3in} 
\onecolumngrid
{\begin{center}\bf \Large{Supplementary material}\end{center} }

\section{Gauge and Fall-Offs}

In this section, we will quickly recap the gauge and metric ansatz introduced in \cite{KP} and then present the more general set of fall-offs that allows the BBMS algebra as a symmetry.

The special double null gauge in 3+1 dimensions is defined by the four conditions \cite{KP}:
\begin{eqnarray}
g^{uu}=0=g^{vv}, \ \ g^{uA}=g^{vA}
\end{eqnarray}
where $u$ and $v$ will eventually be identified as non-compact null coordinates, and $A$ is a direction on the sphere. These lead to the following exact Killing vector equations which will be useful to us:
\begin{eqnarray}\label{exactKilling}
        \mathcal{L}_{\xi} g^{uu} = 0, \
        \mathcal{L}_{\xi} g^{vv} = 0, \ 
        \mathcal{L}_{\xi} g^{uA} = \mathcal{L}_{\xi} g^{vA}
\end{eqnarray}

A general metric in this gauge can always be written as follows \cite{KP}:
\beq \label{doublenull}
ds^2=-e^{\lambda}du\ dv +\Big(\frac{v-u}{2}\Big)^2\Omega_{AB}(dx^A-\alpha^A du-\alpha^A dv)(dx^B-\alpha^B du-\alpha^B dv)
\eeq 
In this paper, we will consider the following set of power law fall-offs for asymptotically flat spacetime, that are more general than those considered in \cite{KP}:
\begin{subequations}\label{polyfalloffs}
    \begin{align}
        \label{polylambda}
        \lambda(u,v,z,\zbr) &= \frac{\lambda_{2}(u,z,\zbr)}{v^2}+\frac{\lambda_{3}(u,z,\zbr)}{v^3}+\frac{\lambda_{4}(u,z,\zbr)}{v^4}+O\big(v^{-5}\big) \\
        \label{polyOmegazz}
        \Omega_{zz}(u,v,z,\zbr) &= \gamma_{zz}(z,\zbr)+\frac{\mathcal{C}_{zz}(u,z,\zbr)}{v}+\frac{\mathcal{D}_{zz}(u,z,\zbr)}{v^2}+\frac{\mathcal{E}_{zz}(u,z,\zbr)}{v^3}+O\big(v^{-4}\big)\\
        \label{polyOmegazw}
        \Omega_{z\zbr}(u,v,z,\zbr) &= \gamma_{z\zbr}(z,\zbr)+\frac{\mathcal{C}_{z\zbr}(u,z,\zbr)}{v}+\frac{\mathcal{D}_{z\zbr}(u,z,\zbr)}{v^2}+\frac{\mathcal{E}_{z\zbr}(u,z,\zbr)}{v^3}+O\big(v^{-4}\big)\\
        \label{polyOmegaww}
        \Omega_{\zbr\zbr}(u,v,z,\zbr) &= \gamma_{\zbr\zbr}(z,\zbr)+\frac{\mathcal{C}_{\zbr\zbr}(u,z,\zbr)}{v}+\frac{\mathcal{D}_{\zbr\zbr}(u,z,\zbr)}{v^2}+\frac{\mathcal{E}_{\zbr\zbr}(u,z,\zbr)}{v^3}+O\big(v^{-4}\big)\\
        \label{polyalphaz}
        \alpha^z(u,v,z,\zbr) &= \frac{\alpha^{z}_{\ 3}(u,z,\zbr)}{v^3}+\frac{\alpha^{z}_{\ 4}(u,z,\zbr)}{v^4}+\frac{\alpha^{z}_{\ 5}(u,z,\zbr)}{v^5}+O\big(v^{-6}\big)\\
        \label{polyalphaw}
        \alpha^{\zbr}(u,v,z,\zbr) &= \frac{\alpha^{\zbr}_{\ 3}(u,z,\zbr)}{v^3}+\frac{\alpha^{\zbr}_{\ 4}(u,z,\zbr)}{v^4}+\frac{\alpha^{\zbr}_{\ 5}(u,z,\zbr)}{v^5}+O\big(v^{-6}\big)
    \end{align}
\end{subequations}
where as in \cite{KP}, the unit-2 sphere metric $\gamma_{AB}$ appearing in \eqref{polyOmegazz}-\eqref{polyOmegaww} is defined as
\beq \gamma_{zz}(z,\zbr)=\gamma_{\zbr\zbr}(z,\zbr)=0, \quad \gamma_{z\zbr}(z,\zbr)=\frac{2}{(1+z\zbr)^2} \label{complexsphere}\eeq

 This results in the following metric fall-offs:
\begin{subequations}\label{boundarycond}
    \begin{align}
        g_{uu} &= g_{vv}= O\big(v^{-4}\big)\\
        g_{uv} &= -\frac{1}{2}+O\big(v^{-2}\big)\\
        g_{AB} &= \frac{1}{4}\, \gamma_{AB}\, v^2+O(v)\\
        g_{uA} &= g_{vA} = O\big(v^{-1}\big)
    \end{align}
\end{subequations}
These fall-offs are more general than those considered in \cite{KP}, and will allow a more general asymptotic symmetry algebra than BMS, as we will demonstrate in the body of the paper. 

Compared to our BMS discussion in \cite{KP} the differences are subtle, but crucial. We allow $\mathcal{C}_{z\zbr}$ as the $O(1/v)$ term in the $\Omega_{z\zbr}$ fall-off, and we also allow the possibility that the second\footnote{Note that the analogue of the Bondi mass aspect constraint in SDN gauge is second order in $u$ \cite{KP}.} integration ``constant'' arising in $\lambda_2$ and the first integration ``constant'' in $\alpha^A_3$ after the Einstein equations are imposed, are allowed. The relevant Einstein equations are presented in \cite{KP, KPBig}. As noted already, $\mathcal{C}_{z\zbr}$ is $u$-independent after the Einstein constraints are imposed -- in other words, it is also an integration ``constant''. 

Given that Einstein equations force $\lambda_1$ to vanish, this is the most general form of power law fall-offs that asymptote to a leading behavior that is Minkowski, with leading behavior of $\Omega_{AB}$ equal to $\gamma_{AB}$.

\section{Shifting the Diffeomorphisms}


In our AKV expressions \eqref{xidef}, we have defined the hypertranslations and hyperrotations {\em not} directly as the corresponding fall-off coefficient of the $\xi$'s. Instead, we have defined them after  appropriate shifts. In this section, we will present the rationale behind the choice of such shifts. 

The discussion is simplest for the leading hypertranslations $\phi(z,\zbr)$, so let us start there. The shift here is of the form
\begin{equation}\label{phishift}
    \xi^v_{(0)}=\phi+T+\Delta_{\gamma}T
\end{equation}
This shift is necessary for us to obtain the relevant piece of the BBMS algebra
\begin{equation}\label{newphihat}
    \widehat{\phi} =\frac{1}{2}(\psi_1\phi_2-\psi_2\phi_1)+\big(Y_{1}^{A}\partial_A\phi_2-Y_{2}^{A}\partial_A\phi_1\big)
\end{equation}
Without the shift (ie., if $\xi^v_{(0)}\equiv \phi$), we would come to the unpleasant conclusion that even when the $\phi_1$ and $\phi_2$ are vanishing, $\widehat \phi$ does not\footnote{In order to make a statement like that we also need to make an assumption about the action of an AKV on $\phi$ of the form $\delta_\xi \phi=0$.}. This means very loosely that we have not ``diagonalized'' the diffeomorphisms suitably, and one way to view the shifts is as a way to avoid this.  
The choice of shift that we made above is essentially unique. A systematic procedure to determine the form of the shift above is as follows.

The gauge parameter corresponding to the hypertranslations $\phi(z,\zbr)$ is $\mathcal{C}_{z\zbr}$. What this means is that the transformation of $\mathcal{C}_{z\zbr}$ under the action of the AKVs in \eqref{finalxi} depends on $\phi$. We obtain an expression for the change in $\mathcal{C}_{z\zbr}$ by evaluating $\delta_{\xi}g^{z\zbr}=\mathcal{L}_{\xi}g^{z\zbr}$ at $O(v^{-3})$ as
\beq\label{deltaCzw} \delta\mathcal{C}_{z\zbr} = \Big[f\, \partial_u+\mathcal{L}_{Y}-\frac{1}{2}\, \psi\Big]\, \mathcal{C}_{z\zbr}-4\, \partial_z\partial_{\zbr}f+2\, \gamma_{z\zbr}\Big(\xi^v_{(0)}-f-\frac{u}{2}\, \psi\Big) \eeq
where $\mathcal{L}_Y$ is the Lie derivative of $\mathcal{C}_{z\zbr}$ with respect to $Y^A$ defined as
\beq\mathcal{L}_{Y}\mathcal{C}_{z\zbr}=Y^A\partial_A\mathcal{C}_{z\zbr}+\big(\partial_AY^A\big)\mathcal{C}_{z\zbr}\eeq
Recalling that on-shell $\mathcal{C}_{z\zbr}=\mathcal{C}^{(0)}_{z\zbr}(z,\zbr)$ and substituting \eqref{fdef} in \eqref{deltaCzw} 
we obtain
\beq
\delta\mathcal{C}^{(0)}_{z\zbr}=\Big[\mathcal{L}_{Y}-\frac{1}{2}\, \psi\Big]\mathcal{C}^{(0)}_{z\zbr}+2\gamma_{z\zbr}(\xi^v_{(0)}-T-\Delta_{\gamma}T) \label{goldstone}
\eeq
As it stands it is clear that $\xi^v_{(0)}$ mixes with the ordinary supertranslations. We would like to remove this mixing, and interpret $\phi(z,\zbr)$ as the diffeomorphism that causes $\mathcal{C}^{(0)}_{z\zbr}$ to be turned on if it was initially zero. 
This immediately suggests that the correct shift is $\xi^v_{(0)}=\phi+T+\Delta_{\gamma}T$,
which defines the hypertranslations $\phi(z,\zbr)$. 
We will see below that a similar logic generalizes readily to subleading hypertranslations as well as hyperrotations. It can also be checked that this shift forces the vanishing of the hatted $\widehat \phi$ on the left hand side of algebra, when $\phi_1$ and $\phi_2$ are set to zero. This feature also generalizes to the other diffeomorphisms.

The general philosophy for identifying the shifts should be clear from the above discussion. It can be stated in the form of an algorithm as follows. First, we identify the function in the metric corresponding to the diffeomorphism to be shifted -- note that there exists a one-to-one correspondence between the diffeomorphisms and independent metric functions \cite{KP}. Then we need to determine how the metric function transforms under the action of the relevant AKV. In order to do so, we evaluate the equation $\delta_{\xi}g^{\mu\nu}=\mathcal{L}_{\xi}g^{\mu\nu}$ for the relevant metric component to the required polynomial order. Next we look at the $u$-independent terms on both sides of the equation. (In the case of the $\mathcal{C}_{z\zbr}$ above, there was a slight extra simplification because on-shell, there was no $u$-dependence at all.) 
We then demand that the inhomogeneous piece in the variation of this $u$-independent part of the metric function be dependent only on the shifted diffeomorphism. This fixes the shift. 


Even though the philosophy is simple, since it may be new, let us see what this procedure yields in the case of hyperrotations $Z^A(z,\zbr)$. The metric function corresponding to hyperrotations is $\alpha^A_3$. The manner in which $\alpha^A_3$ transforms under the action of AKVs may be obtained by evaluating $\delta_{\xi}g^{uA}=\mathcal{L}_{\xi}g^{uA}$ at $O(v^{-3})$ to obtain
\beq\label{deltaalpha3A}\delta\alpha_{3}^{A} = \big[f\, \partial_u+\mathcal{L}_{Y}+\psi\big]\alpha_3^{A}+2\, \xi_{(2)}^{A}+4\, u\, D^{A}f-2\, \mathcal{C}^{AB}\, D_{B}f \eeq
where
\beq\mathcal{L}_Y\alpha_{3}^{A} = Y^B\, \partial_B \alpha_{3}^{A}- \alpha_{3}^{B}\, \partial_BY^A.\eeq
is the Lie derivative of $\alpha^A_3$ with respect to $Y^A$.
Now, on the solution space (ie., when Einstein equations are satisfied) one can check that \cite{KPBig}
\beq\label{alpha3z}\begin{aligned}\partial_u\alpha_{3}^{z} &= -2\, D_{z}\mathcal{C}^{zz}+2\, D_{\zbr}\mathcal{C}^{z\zbr}\\ \implies \alpha^z_{3}(u,z,\zbr) &= -2\int^u_{-\infty}du'\big(D_{z}\mathcal{C}^{zz}-D_{\zbr}\mathcal{C}^{z\zbr}\big)+a^z(z,\zbr)\end{aligned}\eeq
along with a similar integral equation for $\alpha^{\zbr}_{3}(u,z,\zbr)$.
Next, substituting \eqref{alpha3z}, \eqref{xi2A} and \eqref{fdef} into \eqref{deltaalpha3A} and isolating the $u$-independent terms, we obtain
\beq\label{deltaaz}\delta a^z=\big[\mathcal{L}_Y+\psi\big]a^z+2\, \tilde Z^z-2\, \mathcal{C}^{zz}_{(0)}\, D_zT-2\, \mathcal{C}^{z\zbr}_{(0)}\, D_{\zbr}T-2\, T D_z\mathcal{C}^{zz}_{(0)}+2\, T D_{\zbr}\mathcal{C}^{z\zbr}_{(0)}\eeq
Following the procedure outlined above, the inhomogeneous part of the variation yields the shift: 
\beq\label{Zzshift} \tilde Z^z=Z^z+\mathcal{C}^{zz}_{(0)}\, D_zT+\mathcal{C}^{z\zbr}_{(0)}\, D_{\zbr}T+T D_z\mathcal{C}^{zz}_{(0)}-T D_{\zbr}\mathcal{C}^{z\zbr}_{(0)}\eeq
A similar procedure for $\alpha^{\zbr}_{3}(u,z,\zbr)$, yields an analogous shift for the $\zbr$-component of the hyperrotations: 
\beq\label{Zwshift} \tilde Z^{\zbr}=Z^{\zbr}+\mathcal{C}^{\zbr\zbr}_{(0)}\, D_{\zbr}T+\mathcal{C}^{\zbr z}_{(0)}\, D_{z}T+T D_{\zbr}\mathcal{C}^{\zbr\zbr}_{(0)}-T D_{z}\mathcal{C}^{z\zbr}_{(0)}\eeq

This is a good juncture to mention a third motivation to introduce the shifts. In deriving our BBMS algebra of diffeomorphisms, we need to consider the action of the AKVs on the diffeomorphisms themselves. This is because loosely speaking the algebra arises from the modified bracket of two diffeomorphisms. 
Since the $Z^A$'s are the independent diffeomorphisms in our shifted language, it is natural to demand that 
\beq
\delta_\xi Z^A =0.
\eeq
This leads to 
\beq\label{deltaZA}\begin{aligned}\delta_{\xi}\tilde Z^z &= \big(\delta\mathcal{C}^{zz}_{(0)}\big)\, D_zT+\big(\delta\mathcal{C}^{z\zbr}_{(0)}\big)\, D_{\zbr}T+T \big(D_z\delta\mathcal{C}^{zz}_{(0)}\big)-T \big(D_{\zbr}\delta\mathcal{C}^{z\zbr}_{(0)}\big)\\ \delta_{\xi}\tilde Z^{\zbr} &= \big(\delta\mathcal{C}^{\zbr\zbr}_{(0)}\big)\, D_{\zbr}T+\big(\delta\mathcal{C}^{z\zbr}_{(0)}\big)\, D_zT+T \big(D_{\zbr}\delta\mathcal{C}^{\zbr\zbr}_{(0)}\big)-T \big(D_{z}\delta\mathcal{C}^{z\zbr}_{(0)}\big)\end{aligned}\eeq
In computing the algebra for the shifted hyperrotations $Z^A$, these expressions are crucial for cancelling out certain undesired contributions, and thereby leading to 
the simple form of our final algebra \eqref{allhats}. 
We repeat here the algebra of the shifted hyperrotations $Z^A$:
\beq\widehat{Z}^A = \big(\psi_{1}Z^A_{2}-\psi_{2}Z^A_{1}\big)+\big(Y_1^{B}\partial_B Z^A_2-Y_2^{B}\partial_B Z^A_1\big)+\big(Z^B_1\partial_BY_2^{A}-Z^B_2\partial_BY_1^{A}\big).\eeq

Following a procedure similar to the ones detailed above, one can determine the shift for the subleading hypertranslations $\tau(z,\zbr)$ as well. The metric function coefficient corresponding to the subleading hypertranslation can be viewed as\footnote{One can also view it as $\mathcal{D}_{z\zbr}$, but the two are related via Einstein constraints \cite{KP}.} $\lambda_2$. It can be checked that $\lambda_2$ transforms under the action of AKVs as
\beq\delta\lambda_2=\big[f\partial_u+\mathcal{L}_Y+\psi\big]\lambda_2-\frac{1}{4}\, \alpha^A_3\, \partial_A\psi+\frac{1}{2}\, \partial_u\alpha^A_3\, \partial_Af-\xi^v_{(1)} \label{lambda2var}\eeq
This is obtained by evaluating $\delta_{\xi}g_{uv}=\mathcal{L}_{\xi}g_{uv}$ at $O(v^{-2})$. Note that $\mathcal{L}_{Y}\lambda_2=Y^A\partial_A\lambda_2$ is the Lie derivative of $\lambda_2$ with respect to $Y^A$. 
By demanding the Einstein equations, we obtain the following constraint \cite{KP}
\beq
\label{lambda2eq}\partial^2_u\lambda_2=-\frac{1}{2}\, D_AD_B\mathcal{N}^{AB}-\frac{1}{8}\, \partial_u\mathcal{N}^{AB}\, \mathcal{C}_{AB}
\eeq
This can be written in integral form as \cite{KP, KPBig}
\beq\label{intlambda2}
\lambda_2=\lambda_2^0(z,\zbr)+u\ \lambda_2^1(z,\zbr)+\Lambda_2(u,z,\zbr)
\eeq
where \beq\Lambda_2(u,z,\zbr)=\int_{-\infty}^{u}du'\int_{-\infty}^{u'}du'' ({\rm RHS \ of \ \eqref{lambda2eq}})\eeq 
even though the details \cite{KPBig} of this last expression are not important here. Next, substituting \eqref{intlambda2}, \eqref{alpha3z} (and the analogous equation for $\alpha^{\zbr}(u,z,\zbr)$) and \eqref{fdef} in \eqref{lambda2var}, and isolating the $u$-independent terms, we obtain
\beq
\delta\lambda_2^0=\big[\psi+\mathcal{L}_{Y}\big]\lambda_2^0+ T\, \lambda_2^1-\xi^v_{(1)}-\frac{1}{4}\, a^A\, \partial_A\psi +\big(D_{\zbr}\mathcal{C}^{z\zbr}-D_z\mathcal{C}^{zz}\big)\, D_zT+\big(D_{z}\mathcal{C}^{z\zbr}-D_{\zbr}\mathcal{C}^{\zbr\zbr}\big)\, D_{\zbr}T
\eeq
Once more as before, 
demanding that the inhomogeneous part of this is the independent diffeomorphism leads to
\beq\xi^v_{(1)}=\tau-\frac{1}{4}\, a^A\, \partial_A\psi+\big(D_{\zbr}\mathcal{C}^{z\zbr}-D_z\mathcal{C}^{zz}\big)\, D_zT+\big(D_{z}\mathcal{C}^{z\zbr}-D_{\zbr}\mathcal{C}^{\zbr\zbr}\big)\, D_{\zbr}T\eeq
This leads to the modified algebra we presented earlier:
\beq\widehat{\tau} = (\psi_1\tau_2-\psi_2\tau_1)+\big(Y_1^A\partial_A\tau_2-Y_2^A\partial_A\tau_1\big)\eeq
Note that we have set $\lambda_2^1=0$ in determining the shift in $\xi^v_{(1)}$ because the most general strategy for identifying the shifts is to fix them for the Riemann flat fall-offs. The complete set of constraints obtained from demanding vanishing Riemann curvature will be discussed elsewhere \cite{KPInfty} along with the general philosophy behind the shifts. In the present paper, we have fixed the shifts using somewhat more {\em ad-hoc} approaches --  for our purposes here, it suffices to note that $\lambda_2=\lambda_2^0(z,\zbr)$ in the Riemann flat metric. 

Despite the relative simplicity of the calculations in this section compared to some of our other discussions, we have decided to present them in detail. This is because this type of shift seems pretty novel to our gauge. The calculations of the hatted objects in the BBMS algebra from the Barnich-Troessaert commutators of AKVs is more involved, but they are a straightforward adaptation/extension of previous BMS discussions in the Bondi gauge to higher orders. 

\section{Covariant Surface Charges}\label{covaurface}

We will need the surface charges to decide which of the integration data corresponds to actual physical parameters.   In the covariant phase space formalism, the expression for the surface charges takes the form
\beq \slashed{\delta}\mathcal{Q}_{\xi}[h;g] = \oint_{S^2} \mathbf{k}_{\xi}[h;g] \eeq
where $\mathbf{k}_{\xi}$ is a $2$-form defined on the $2$-sphere at future null infinity. Note that $g_{\mu\nu}$ denotes the background metric and $h_{\mu\nu}=\delta g_{\mu\nu}$ is the infinitesimal perturbation about the background $g_{\mu\nu}$ in the solution space. 
Also note that we have made use of $\slashed{\delta}$ instead of $\delta$ in order to emphasize that the right-hand-side is not an exact differential in the space of metrics. If it is, then the charges are said to be \textit{integrable}.

We will use the expression for $\mathbf{k}_{\xi}$ that corresponds to the  so-called Iyer-Wald \cite{Iyer} charges \footnote{The Barnich-Brandt form \cite{Brandt} of the charges do not affect our qualitative conclusions, only the final expressions of the charges.}:
\beq
\mathbf{k}_{\xi}[h;g]=\frac{\sqrt{-g}}{16\pi G}(d^2x)_{\mu\nu}\Big[\xi^{\mu} \nabla_{\sigma}h^{\nu\sigma}-\xi^{\mu} \nabla^{\nu}h+\xi_{\sigma} \nabla^{\nu}h^{\mu\sigma}+\frac{1}{2} h \nabla^{\nu}\xi^{\mu}-h^{\rho\nu} \nabla_{\rho}\xi^{\mu}-(\mu\leftrightarrow\nu)\Big]
\eeq
where $g=\text{det}{g_{\mu\nu}}$, $h=g^{\mu\nu}h_{\mu\nu}$ and $\nabla_{\mu}$ is the covariant derivative associated with the background metric $g_{\mu\nu}$. $G$ is the 4d Newton's  constant. Note that $g_{\mu\nu}$ is to be used for raising and lowering the Greek indices $\mu,\nu,$ etc. For the asymptotic Killing vectors \eqref{finalxi} and the metric fall-offs \eqref{polyfalloffs} in the double null gauge, the expression for the Iyer-Wald charge may be evaluated as follows:
\beq\begin{aligned}\slashed{\delta}\mathcal{Q}_{\xi}[h;g] &= \frac{1}{16\pi G}\lim_{v\to\infty}\int d^2\omega\, \frac{1}{2} e^{\lambda} \Big(\frac{v-u}{2}\Big)^2 (-\text{det}\Omega_{AB})^{1/2}\, \Big[\xi^v \big(\nabla^uh-\nabla_{\sigma}h^{u\sigma}+\nabla^vh^{u}_{\ v}-\nabla^{u}h^{v}_{\ v}\big) \\& -\xi^{u}\big(\nabla^{v}h-\nabla_{\sigma}h^{v\sigma}-\nabla^{v}h^{u}_{\ u}+\nabla^{u}h^{v}_{\ u}\big)+\xi^A\big(\nabla^{v}h^{u}_{\ A}-\nabla^{u}h^{v}_{\ A}\big)+\frac{1}{2}h\big(\nabla^v\xi^{u}-\nabla^u\xi^{v}\big) \\& -h^{v\sigma}\nabla_{\sigma}\xi^{u}+h^{u\sigma}\nabla_{\sigma}\xi^v\Big]\end{aligned}\eeq 
The terms here can be explicitly computed to be
\beq
\nabla^uh-\nabla_{\sigma}h^{u\sigma}+\nabla^vh^{u}_{\ v}-\nabla^{u}h^{v}_{\ v} = -4\, \delta\lambda_2\,\frac{1}{v^3}+O\big(v^{-4}\big)    
\eeq
\beq\begin{aligned}
\nabla^{v}h-\nabla_{\sigma}h^{v\sigma}-\nabla^{v}h^{u}_{\ u}+\nabla^{u}h^{v}_{\ u} &= \big(8\, \partial_u\delta\lambda_2+4\, D_zD_z\delta\mathcal{C}^{zz}+4\, D_{\zbr}D_{\zbr}\delta\mathcal{C}^{\zbr\zbr}\\&-8\, D_zD_{\zbr}\mathcal{C}^{z\zbr}+\mathcal{C}^{zz}\, \delta\mathcal{N}_{zz}+\mathcal{C}^{\zbr\zbr}\, \delta\mathcal{N}_{\zbr\zbr}\big)\, \frac{1}{v^2}+O\big(v^{-3}\big)    
\end{aligned}\eeq
\beq\begin{aligned}
\nabla^{v}h^{u}_{\ A}-\nabla^{u}h^{v}_{\ A} &= \Big(-\gamma_{AB}\, \partial_u\delta\alpha_{3}^{B}\Big)\, \frac{1}{v}+\Big(-\delta\mathcal{C}_{AB}\, \partial_u\alpha_{3}^{B}-\mathcal{C}_{AB}\, \partial_u\delta\alpha_{3}^{B}-\frac{1}{2}\, \mathcal{N}_{AB}\, \delta\alpha_{3}^{B} \\& -\gamma_{AB}\, \big(\delta\alpha_{3}^{B}-2\, u\, \partial_u\delta\alpha_{3}^{B}+\partial_u\delta\alpha_{4}^{B}\big)\Big)\, \frac{1}{v^2}+O\big(v^{-3}\big)    
\end{aligned}\eeq
\beq\begin{aligned}
\nabla^v\xi^{u}-\nabla^u\xi^{v} &= -2\, \psi-\gamma_{AB}\, Y^A\, \partial_u\alpha_{3}^{B}\, \frac{1}{v}+O\big(v^{-2}\big) 
\end{aligned}\eeq
\beq\begin{aligned}
-h^{v\sigma}\nabla_{\sigma}\xi^{u}+h^{u\sigma}\nabla_{\sigma}\xi^v &= \frac{1}{2}\, \Big(4\, \delta\lambda_2\, \psi-2\, \delta\alpha_3^{A}\, D_{A}\psi-4\, \gamma_{AB}\, Y^{A}\, \delta\alpha_{3}^{B}+\mathcal{N}_{AB}\, Y^{A}\, \delta\alpha^{B}\Big)\, \frac{1}{v^2}+O\big(v^{-3}\big) 
\end{aligned}\eeq
\beq\begin{aligned}
\frac{1}{2} e^{\lambda} \Big(\frac{v-u}{2}\Big)^2 (-\text{det}\Omega_{AB})^{1/2} &= \frac{1}{8}\, \gamma_{z\zbr}\, v^2+\frac{1}{8}\, (\mathcal{C}_{z\zbr}-2\, u\, \gamma_{z\zbr})\, v \\& +\frac{1}{16} \Big(2\, \gamma_{z\zbr}\, u^2+2\, \lambda_2\, \gamma_{z\zbr}+2\, \mathcal{D}_{z\zbr}-4\, u\, \mathcal{C}_{z\zbr}-\gamma^{z\zbr}\, \mathcal{C}_{zz}\, \mathcal{C}_{\zbr\zbr}\Big)+O\big(v^{-1}\big)
\end{aligned}\eeq
Putting everything together, we get
\beq\label{doublenullcharge1}
\slashed{\delta}\mathcal{Q}_{\xi}[h;g] = \frac{1}{16\pi G}\lim_{v\to\infty}\int dz d\zbr\, \bigg[
\frac{1}{4}\, \Big(Y^z\big(D_{\zbr}\delta\mathcal{C}_{zz}-D_z\delta\mathcal{C}_{z\zbr}\big)+Y^{\zbr}\big(D_{z}\delta\mathcal{C}_{\zbr\zbr}-D_{\zbr}\delta\mathcal{C}_{z\zbr}\big)-\psi\, \delta\mathcal{C}_{z\zbr}\Big)\, v+O\big(v^{0}\big)\bigg]\eeq
To show the finiteness of the charges we need to establish that the $O\big(v\big)$ term vanishes. The terms in the parenthesis at $O\big(v\big)$ in the above expression can be rewritten as
\beq\begin{aligned}
    Y^z\big(D_{\zbr}\delta\mathcal{C}_{zz}-D_z\delta\mathcal{C}_{z\zbr}\big) &+Y^{\zbr}\big(D_{z}\delta\mathcal{C}_{\zbr\zbr}-D_{\zbr}\delta\mathcal{C}_{z\zbr}\big)-\psi\, \delta\mathcal{C}_{z\zbr} \\ &= Y^z\, D_{\zbr}\delta\mathcal{C}_{zz}+Y^{\zbr}\, D_{z}\delta\mathcal{C}_{\zbr\zbr}-Y^{z}D_z\delta\mathcal{C}_{z\zbr}-Y^{\zbr}D_{\zbr}\delta\mathcal{C}_{z\zbr}-\big(D_zY^z+D_{\zbr}Y^{\zbr}\big)\delta\mathcal{C}_{z\zbr}\\
    &=D_{\zbr}\big(Y^z\, \delta\mathcal{C}_{zz}\big)+D_{z}\big(Y^{\zbr}\, \delta\mathcal{C}_{\zbr\zbr}\big)-D_z\big(Y^z\, \delta\mathcal{C}_{z\zbr}\big)-D_{\zbr}\big(Y^{\zbr}\, \delta\mathcal{C}_{z\zbr}\big)\\ &=D_z\big(Y^{\zbr}\, \delta\mathcal{C}_{\zbr\zbr}-Y^z\, \delta\mathcal{C}_{z\zbr}\big)+D_{\zbr}\big(Y^z\, \delta\mathcal{C}_{zz}-Y^{\zbr}\, \delta\mathcal{C}_{z\zbr}\big)
\end{aligned}\eeq
These ``total" derivative terms vanish when we do the integration over the 2-sphere. Since there are no $O\big(v\big)$ terms, the surface charges are guaranteed to remain finite in the limit $v\rightarrow\infty$. Thus on taking the limit, only the $O\big(v^0\big)$ terms remain and can be evaluated to be
\beq\begin{aligned}
\slashed{\delta}\mathcal{Q}_{\xi}[h;g] &= -\frac{1}{16\pi G}\int dz d\zbr\, \Big(D_z\big(Y^{\zbr}\, \delta\mathcal{C}_{\zbr\zbr}-Y^z\, \delta\mathcal{C}_{z\zbr}\big)+D_{\zbr}\big(Y^z\, \delta\mathcal{C}_{zz}-Y^{\zbr}\, \delta\mathcal{C}_{z\zbr}\big)\Big)\, u\\&+ \frac{1}{16\pi G}\int d^2\Omega\, \Big[\frac{3}{4}\, \psi\, \delta\lambda_2-f\, \partial_u\delta\lambda_2-\frac{1}{8}\, Y_A\, \partial_u\delta\alpha^A_4-\frac{3}{8}\, Y_A\, \delta\alpha^A_{3}-\frac{1}{8}f\, \mathcal{C}^{AB}\, \delta\mathcal{N}_{AB}\\&+\frac{1}{16}\psi\, \mathcal{C}^{zz}\, \delta\mathcal{C}_{zz}+\frac{1}{16}\psi\, \mathcal{C}^{\zbr\zbr}\, \delta\mathcal{C}_{\zbr\zbr}-\frac{1}{8}\psi\, \mathcal{C}^{z\zbr}\, \delta\mathcal{C}_{z\zbr}-\frac{3}{4}u\, \psi\, \gamma^{z\zbr}\delta\mathcal{C}_{z\zbr}\\&+\frac{1}{4}\, Y^{\zbr}\, \mathcal{C}^{zz}\, D_{\zbr}\delta\mathcal{C}_{zz}+\frac{1}{4}\, Y^{z}\, \mathcal{C}^{\zbr\zbr}\, D_{z}\delta\mathcal{C}_{\zbr\zbr}-\frac{1}{4}\, Y^{\zbr}\, \mathcal{C}^{zz}\, D_{z}\delta\mathcal{C}_{z\zbr}-\frac{1}{4}\, Y^{z}\, \mathcal{C}^{\zbr\zbr}\, D_{\zbr}\delta\mathcal{C}_{z\zbr}\\&+\frac{1}{2}\, Y^{z}\, \mathcal{C}^{z\zbr}\, D_{\zbr}\delta\mathcal{C}_{zz}+\frac{1}{2}\, Y^{\zbr}\, \mathcal{C}^{z\zbr}\, D_{z}\delta\mathcal{C}_{\zbr\zbr}-\frac{1}{2}\, Y^{z}\, \mathcal{C}^{z\zbr}\, D_{z}\delta\mathcal{C}_{z\zbr}-\frac{1}{2}\, Y^{\zbr}\, \mathcal{C}^{z\zbr}\, D_{\zbr}\delta\mathcal{C}_{z\zbr}\\&+\frac{1}{4}\, Y^{\zbr}\, \delta\mathcal{C}^{zz}\, D_{\zbr}\mathcal{C}_{zz}+\frac{1}{4}\, Y^{z}\, \delta\mathcal{C}^{\zbr\zbr}\, D_{z}\mathcal{C}_{\zbr\zbr}-\frac{1}{4}\, Y^{\zbr}\, \delta\mathcal{C}^{zz}\, D_{z}\mathcal{C}_{z\zbr}-\frac{1}{4}\, Y^{z}\, \delta\mathcal{C}^{\zbr\zbr}\, D_{\zbr}\mathcal{C}_{z\zbr}\\&+\frac{1}{2}\, Y^{z}\, \delta\mathcal{C}^{z\zbr}\, D_{\zbr}\mathcal{C}_{zz}+\frac{1}{2}\, Y^{\zbr}\, \delta\mathcal{C}^{z\zbr}\, D_{z}\mathcal{C}_{\zbr\zbr}-\frac{1}{2}\, Y^{z}\, \delta\mathcal{C}^{z\zbr}\, D_{z}\mathcal{C}_{z\zbr}-\frac{1}{2}\, Y^{\zbr}\, \delta\mathcal{C}^{z\zbr}\, D_{\zbr}\mathcal{C}_{z\zbr}\\&-\frac{1}{8}\, D_A\big(\psi\, \delta\alpha^A_3\big)+\frac{1}{2}\, D_z\big(f\, D_{\zbr}\delta\mathcal{C}^{z\zbr}\big)+\frac{1}{2}\, D_{\zbr}\big(f\, D_{z}\delta\mathcal{C}^{z\zbr}\big)-\frac{1}{2}\, D_z\big(f\, D_{z}\delta\mathcal{C}^{zz}\big)\\&-\frac{1}{2}\, D_{\zbr}\big(f\, D_{\zbr}\delta\mathcal{C}^{\zbr\zbr}\big)\Big]
\end{aligned}\eeq
where $\int d^2\Omega=\int dzd\zbr\sqrt{\gamma}$. Once again, the terms in the first integral and the terms in the last two lines of the second integral have been written in the form of total derivatives and hence vanish upon integrating over the 2-sphere. Thus the expression for the surface charges can be written as
\beq\begin{aligned}
\slashed{\delta}\mathcal{Q}_{\xi}[h;g] &=\frac{1}{16\pi G}\int d^2\Omega\, \Big[\frac{3}{4}\, \psi\, \delta\lambda_2-f\, \partial_u\delta\lambda_2-\frac{1}{8}\, Y_A\, \partial_u\delta\alpha^A_4-\frac{3}{8}\, Y_A\, \delta\alpha^A_{3}-\frac{1}{8}f\, \mathcal{C}^{AB}\, \delta\mathcal{N}_{AB}\\&+\frac{1}{16}\psi\, \mathcal{C}^{zz}\, \delta\mathcal{C}_{zz}+\frac{1}{16}\psi\, \mathcal{C}^{\zbr\zbr}\, \delta\mathcal{C}_{\zbr\zbr}-\frac{1}{8}\psi\, \mathcal{C}^{z\zbr}\, \delta\mathcal{C}_{z\zbr}-\frac{3}{4}u\, \psi\, \gamma^{z\zbr}\delta\mathcal{C}_{z\zbr}\\&+\frac{1}{4}\, Y^{\zbr}\, \mathcal{C}^{zz}\, D_{\zbr}\delta\mathcal{C}_{zz}+\frac{1}{4}\, Y^{z}\, \mathcal{C}^{\zbr\zbr}\, D_{z}\delta\mathcal{C}_{\zbr\zbr}-\frac{1}{4}\, Y^{\zbr}\, \mathcal{C}^{zz}\, D_{z}\delta\mathcal{C}_{z\zbr}-\frac{1}{4}\, Y^{z}\, \mathcal{C}^{\zbr\zbr}\, D_{\zbr}\delta\mathcal{C}_{z\zbr}\\&+\frac{1}{2}\, Y^{z}\, \mathcal{C}^{z\zbr}\, D_{\zbr}\delta\mathcal{C}_{zz}+\frac{1}{2}\, Y^{\zbr}\, \mathcal{C}^{z\zbr}\, D_{z}\delta\mathcal{C}_{\zbr\zbr}-\frac{1}{2}\, Y^{z}\, \mathcal{C}^{z\zbr}\, D_{z}\delta\mathcal{C}_{z\zbr}-\frac{1}{2}\, Y^{\zbr}\, \mathcal{C}^{z\zbr}\, D_{\zbr}\delta\mathcal{C}_{z\zbr}\\&+\frac{1}{4}\, Y^{\zbr}\, \delta\mathcal{C}^{zz}\, D_{\zbr}\mathcal{C}_{zz}+\frac{1}{4}\, Y^{z}\, \delta\mathcal{C}^{\zbr\zbr}\, D_{z}\mathcal{C}_{\zbr\zbr}-\frac{1}{4}\, Y^{\zbr}\, \delta\mathcal{C}^{zz}\, D_{z}\mathcal{C}_{z\zbr}-\frac{1}{4}\, Y^{z}\, \delta\mathcal{C}^{\zbr\zbr}\, D_{\zbr}\mathcal{C}_{z\zbr}\\&+\frac{1}{2}\, Y^{z}\, \delta\mathcal{C}^{z\zbr}\, D_{\zbr}\mathcal{C}_{zz}+\frac{1}{2}\, Y^{\zbr}\, \delta\mathcal{C}^{z\zbr}\, D_{z}\mathcal{C}_{\zbr\zbr}-\frac{1}{2}\, Y^{z}\, \delta\mathcal{C}^{z\zbr}\, D_{z}\mathcal{C}_{z\zbr}-\frac{1}{2}\, Y^{\zbr}\, \delta\mathcal{C}^{z\zbr}\, D_{\zbr}\mathcal{C}_{z\zbr}\Big]
\end{aligned}\eeq
Note that the variation $\delta g_{\mu\nu}$ is really the difference between the background metric $g_{\mu\nu}$ and the metric $g'_{\mu\nu}$ obtained after the vriation. 
It will commute with the partial derivative $\partial_u$ and the covariant derivative $D_A$ to the linear order in variations we are dealing with. Similar manipulations were used in the Bondi gauge in \cite{BMSChargeAlgebra}. 
This commuting property has also played a vital role in re-writing certain terms at  $O(v)$ in \eqref{doublenullcharge1} as well:
\beq\begin{aligned}
-\frac{1}{8}\, (\gamma_{z\zbr})^2\, \Big[Y^z\, \partial_u\big(\delta\alpha_{3}^{\zbr}\big)+Y^{\zbr}\, \partial_u\big(\delta\alpha_{3}^{z}\big)\Big] &= -\frac{1}{8}\, (\gamma_{z\zbr})^2\, \Big[Y^z\, \delta\big(\partial_u\alpha_{3}^{\zbr}\big)+Y^{\zbr}\, \delta\big(\partial_u\alpha_{3}^{z}\big)\Big] \\&= \frac{1}{4}\Big[Y^z\, \delta\big(D_{\zbr}\mathcal{C}^{\zbr\zbr}-D_{z}\mathcal{C}^{z\zbr}\big)+Y^{\zbr}\, \delta\big(D_{z}\mathcal{C}^{zz}-D_{\zbr}\mathcal{C}^{z\zbr}\big)\Big] \\&=  \frac{1}{4}\Big[Y^z\big(D_{\zbr}\delta\mathcal{C}_{zz}-D_z\delta\mathcal{C}_{z\zbr}\big)+Y^{\zbr}\big(D_{z}\delta\mathcal{C}_{\zbr\zbr}-D_{\zbr}\delta\mathcal{C}_{z\zbr}\big)\Big]
\end{aligned}\eeq
In any event, using these facts, the above expression for the surface charge can be written in the final form
\beq\begin{aligned}
\slashed{\delta}\mathcal{Q}_{\xi}[h;g] &=\frac{1}{16\pi G}\delta\int d^2\Omega\, \Big[\frac{3}{4}\, \psi\, \lambda_2-f\, \partial_u\lambda_2-\frac{1}{8}\, Y_A\, \partial_u\alpha^A_4-\frac{3}{8}\, Y_A\, \alpha^A_{3}\\&+\frac{1}{32}\psi\, \mathcal{C}^{zz}\, \mathcal{C}_{zz}+\frac{1}{32}\psi\, \mathcal{C}^{\zbr\zbr}\, \mathcal{C}_{\zbr\zbr}-\frac{1}{16}\psi\, \mathcal{C}^{z\zbr}\, \mathcal{C}_{z\zbr}-\frac{3}{4}u\, \psi\, \gamma^{z\zbr}\mathcal{C}_{z\zbr}\\&+\frac{1}{4}\, Y^{\zbr}\, \mathcal{C}^{zz}\, D_{\zbr}\mathcal{C}_{zz}+\frac{1}{4}\, Y^{z}\, \mathcal{C}^{\zbr\zbr}\, D_{z}\mathcal{C}_{\zbr\zbr}-\frac{1}{4}\, Y^{\zbr}\, \mathcal{C}^{zz}\, D_{z}\mathcal{C}_{z\zbr}-\frac{1}{4}\, Y^{z}\, \mathcal{C}^{\zbr\zbr}\, D_{\zbr}\mathcal{C}_{z\zbr}\\&+\frac{1}{2}\, Y^{z}\, \mathcal{C}^{z\zbr}\, D_{\zbr}\mathcal{C}_{zz}+\frac{1}{2}\, Y^{\zbr}\, \mathcal{C}^{z\zbr}\, D_{z}\mathcal{C}_{\zbr\zbr}-\frac{1}{2}\, Y^{z}\, \mathcal{C}^{z\zbr}\, D_{z}\mathcal{C}_{z\zbr}-\frac{1}{2}\, Y^{\zbr}\, \mathcal{C}^{z\zbr}\, D_{\zbr}\mathcal{C}_{z\zbr}\Big]\\&-\frac{1}{16\pi G}\int d^2\Omega\, \Big[\frac{f}{8}\, \mathcal{C}^{AB}\, \partial_u\delta\mathcal{C}_{AB}\Big]
\end{aligned}\eeq

To conclude, the final charge expression above demonstrates not just that the charges are finite and non-vanishing -- it also shows that the relevant metric functions corresponding to the BBMS diffeomorphisms are indeed present in them. In particular, note that $\lambda_2$ and $\alpha^A_3$ (and not just their $u$-derivatives) appear, as well as $\mathcal{C}_{z\zbr}$. But let us also note a difference between BMS charges and the charges we have found here. In the BMS case, the BMS diffeomorphisms (eg., supertranslations) as well as the metric parameters that are directly or indirectly affected by those diffeomorphisms (eg., shear, mass aspect) appear in the charges. This still remains true in our gauge. But the {\em extra} BBMS diffeomorphisms do not appear in the charge expressions above, only the BMS diffeomorphisms do. Despite this, the metric parameters that can be varied by the action of the extra diffeomorphisms do show up in the charges -- $\lambda_2$ and $\alpha^A_3$ (and not just their $u$-derivatives) appear, as well as $\mathcal{C}_{z\zbr}$. This suggests that the status of the new BBMs diffeomorphisms is somewhere in between trivial gauge transformations and global symmetries. The shear $\mathcal{C}_{z\zbr}$ is quite naturally identified (see \eqref{goldstone}) as a Goldstone boson of broken hypertranslations in the same sense that ordinary shear is a Goldstone mode of broken supertranslations. But unlike the ordinary supertranslations, there is no quantity akin to the mass aspect for hypertranslations. We call such transformations {\em marginal} symmetries to distinguish them from trivial and non-trivial asymptotic diffeomorphisms. It is clearly of interest to understand them better.

\end{document}